# A biomechanical breast model evaluated with respect to MRI data collected in three different positions

Anna Mîra[a,b]*, Ann-Katherine Carton[b], Serge Muller[b] and Yohan Payan[a]

[a]Univ. Grenoble Alpes, CNRS, Grenoble INP, VetAgro Sup, TIMC-IMAG, 38000 Grenoble, France

[b] GE Healthcare, 78530 Buc, France

*Corresponding author: anna.mira@univ-grenoble-alpes.fr.

## Abstract

*Background:* Mammography is a specific type of breast imaging that uses low-dose X-rays to detect cancer in early stage. During the exam, the women breast is compressed between two plates in order to even out the breast thickness and to spread out the soft tissues. This technique improves exam quality but can be uncomfortable for the patient. The perceived discomfort can be assessed by the means of a breast biomechanical model. Alternative breast compression techniques may be computationally investigated trough finite elements simulations.

*Methods:* The aim of this work is to develop and evaluate a new biomechanical Finite Element (FE) breast model. The complex breast anatomy is considered including adipose and glandular tissues, muscle, skin, suspensory ligaments and pectoral fascias. Material hyper-elasticity is modeled using the Neo-Hookean material models. The stress-free breast geometry and subject-specific constitutive models are derived using tissues deformations measurements from MR images.

*Findings:* The breast geometry in three breast configurations were computed using the breast stress-free geometry together with the estimated set of equivalent Young's modulus ($E_{breast}^r = 0.3\ kPa, E_{breast}^l = 0.2\ kPa, E_{skin} = 4\ kPa, E_{fascia} = 120\ kPa$). The Hausdorff distance between estimated and measured breast geometries for prone, supine and supine tilted configurations is equal to 2.17 mm, 1.72 mm and 5.90 mm respectively.

*Interpretation:* A subject-specific breast model allows a better characterization of breast mechanics. However, the model presents some limitations when estimating the supine tilted breast configuration. The results show clearly the difficulties to characterize soft tissues mechanics at large strain ranges with Neo-Hookean material models.

Keywords 6: breast, breast support matrix, finite elements, hyper-elasticity, stress-free geometry, optimization.

# Introduction

## Clinical Background

Today, mammography is the primary imaging modality for breast cancer screening and plays an important role in cancer diagnosis. Subtle soft-tissue findings and micro-calcifications that may represent early breast cancer are visualized by means of X-rays images. After investigation, the abnormal findings are taken in charge for further evaluation.

A standard mammographic protocol always includes breast compression prior to image acquisition. Women breast is compressed between two plates until a nearly uniform breast thickness is obtained. The breast flattening improves diagnostic image quality[1] and reduces the absorbed dose of ionizing photons[2]. However, the discomfort and pain produced by this procedure sometimes might deter women from attending breast screening by mammography [3]. Fleming et al[4] show in a study of 2500 women that 15% of those who skipped the second appointment cited an unpleasant or painful first mammogram.

Nowadays, the European Commission recommends a force standardized breast compression, i.e. the compression stops at a level of force just below the subject's pain threshold or to the maximum setting of the machine (not to exceed 200 N). Some research[5] indicates that with a reduced level of compression (10N vs 30N), 24% of women did not experience a difference in breast thickness. If breast thickness is not reduced when compression force is applied, then discomfort is increased with no benefit in image quality.

An important improvement concerning the patient comfort could be achieved with the emergence of Full-Field Digital Mammography (FFDM). Several studies have shown that digital mammography is better in terms of image quality [6,7] and radiation dose [2,8] than Film–Screen Mammography (FSM). Therefore, there is an opportunity to leverage the potential of the recent imaging technologies to investigate alternative breast compression techniques, considering the patient comfort in addition to an improved image quality and a reduced ionizing radiation dose. The aim of this work is to develop and evaluate a biomechanical Finite Element (FE) breast model using in-vivo measured deformations due to gravity loading. In a near future, this model will be calibrated and used to investigate alternative breast compression strategies.

## Related work

Biomechanical modelling of breast tissues is widely investigated for various medical applications such as surgical procedure training, pre-operative planning, diagnosis and clinical biopsy, image guided surgery, image registration, and material parameter estimation. For the last 20 years, several research groups have presented their breast models based on finite elements theory. The complexity and relevance to breast anatomy of each model depend on the research purpose for which it was designed.

Several groups have proposed biomechanical breast models to register uncompressed volumetric breast data to the compressed one[9–14], or to compressed projection mammographic data[13]. Within this framework, the authors modeled the breast deformation from prone to compressed prone position assuming linear elastic materials, zero residual stress and Dirichlet boundary conditions.

However, compression-like breast deformation is too limited to characterize global breast mechanics. Applications such as image guided surgery or preoperative planning imply a wider range of deformations. Therefore biomechanical breast models capable of estimating breast deformation between supine and prone positions (named multi loading simulations) were proposed[15–19]. Considering the involved large deformation, these models need to be more accurate with respect to mechanical and anatomical breast properties. In this respect, a patient-specific model is needed considering more personalized boundary conditions, material models and a better representation of breast anatomy.

Most of proposed patient-specific models are using volumetric data from MR images [13,16,20] or CT images [14,21] to compute the 3D breast geometry. Acquired data represents deformed breast soft tissues due to in-vivo conditions, and therefore includes initial pre-stresses. The tissues pre-stresses are difficult to measure in in-vivo conditions, therefore a breast stress-free configuration must be estimated. The work by Eiben and colleagues[19] compared three numerical methods to estimating breast stress-free geometry: 1) the approximation where only the gravity direction is inverted; 2) the approximation using an inverse finite deformation approach [22]; 3) the approximation using an iterative fixed-point type algorithm [23]. Authors showed that, in multi-gravity loading simulations context, a fixed-point based iterative algorithm provides the best trade-off between involved computation resources and resulting accuracy.

Factors that are likely to affect the biomechanical models' accuracy, such as mesh density, FE solver, materials models and boundary conditions, were analyzed by Tanner et al[24]. According to the authors, a finest definition of model boundary conditions will considerably improve the resultant tissues deformation. It was also shown by Carter and colleagues[25] that imposing Dirichlet conditions on the chest wall results in an over constrained model underestimating the breast lateral displacement. Therefore, the most recent models include a frictionless surface between chest wall and breast [18,25,26].

Accuracy of various constitutive models to estimate tissues behavior was compared within a multi-gravity loading framework[27]. The authors showed that, under gravity loading, the resulting strain exceeds the linear domain. Thus, a hyper-elastic model is mandatory to describe the reported range of in-vivo displacements. Mechanical behavior of breast tissues is usually modeled using Neo-Hookean materials. Even if nearly all studies converged on the same constitutive model, there are important variation between the estimated values of equivalent Young's modulus, ranging between 0.2 and 60 kPa. Such large variability can be explained by inter-individual diversity[28] but also by the difference in employed numerical method or experimental set-ups. The main technics for Young's modulus identification fall into two categories: one based on ex-vivo indentation tests and the second based on in-vivo measurements coupled with finite element modeling. The latter combines nonlinear optimization techniques with biomechanical models to register multi-gravity loading deformations (e.g., prone to supine deformation) [9,16,20,30–31]. The authors report very low Young's modulus values of breast tissue compared to the ones provided by an indentation ex-vivo method [32,33] (0.2-6 kPa compared to 3-60 kPa). Therefore, the estimation of patient-specific constitutive parameters of the Neo-Hookean model will also be addressed in this work.

This paper focuses on patient-specific data to compute breast geometry and to estimate individual mechanical tissues properties. We believe indeed such models are the only capable of keeping the simulation fidelity to the real breast deformations. The scope of the present work is 1) to create a subject-specific biomechanical model capable of simulating global breast deformations due to gravity loads and, 2) to evaluate the accuracy of this biomechanical model using additional MR data that were not included in the calibration process. To be as realistic as possible, our model will consider breast heterogeneity, anisotropy, sliding boundary conditions, initial pre-stresses and personalized hyper-elastic properties of breast tissue. In addition, new types of soft tissue will be included representing the breast support matrix composed of suspensory ligaments and fascias. In addition, among all existing models that were designed from patient MRI data, none have been quantitatively evaluated, to our knowledge, on an additional data set of the same patient. Our model was built using prone and supine breast configurations collected on MRI data of a volunteer and was evaluated in supine tilted configuration (~ 45 degrees) of the same volunteer.

## Methods
### MR images of the breast
Patient-specific data was acquired using MRI modality, enabling in-vivo deformation measurements in a large field of view. The images were acquired with a Siemens 3T MRI scanner with T2 weighted image sequences. The in-plane image resolution was 0.5x0.5 mm, and the slice thickness was 0.6 mm. During this acquisition, the contact between the breasts and the contours of the MRI tunnel, or with the patient body (arms, thorax), was minimized.

The volunteer participating to this study agreed to participate in an experiment part of a pilot study approved by an ethical committee (MammoBio MAP-VS pilot study). The volunteer is 59 years old and has a A-cup breast size. Three different positioning configurations were considered: prone, supine and supine titled (~ 45 deg). These positions were chosen to assess the largest possible deformations with minimal contact areas between the volunteer and the relatively narrow MRI scanner tunnel.

Breast tissue are known to be extremely soft; the breast volume is therefore subject to large deformations during the re-positioning of the volunteer. In some breast surrounding zones, such as the sternal segment or the inframammary fold , the presence of stiff fibrous tissue fixes the breast soft tissues to the thoracic rib cage[34] thus limiting the elastic deformations. These areas were chosen to be the support of 4 fiducials skin markers used latter to assess rigid body motions.

### Image pre-processing

During the imaging acquisition process, the volunteer was moved in and out the MRI scanner. Therefore, the breast not only undergone an elastic transformation, but also a rigid one. The rigid image transform was subtracted by aligning the chest wall line from prone and supine tilted positions to the supine one. The rigid registration was implemented using the gradient descent-based algorithm minimizing the image cross correlation (VTK library). To speed-up the registration process, the rigid transform was initialized with the transform computed from fiducials marker's positions.

Thereafter, MR images (Fig. 1.a) were segmented using a semi-automated active contour method implemented in ITK-snap software[35] (Fig. 1.b). Assuming the adipose and glandular tissue have a similar mechanical stiffness, the volume of interest was classified in 2 types of tissue: muscle and breast tissue. The skin and the Cooper's ligaments are usually not visible in the MR images; therefore, they were considered as components of breast tissue.

### 3D Geometry and Finite Element mesh

Following image segmentation, two surface meshes were created. The first one represents the contours of breast tissue and the second one represents the thoracic cage and muscle contours (Fig. 1.b). The surface meshes were imported with SpaceClaim Direct Modeler[36] and converted to NURB surfaces (Fig. 1.c).

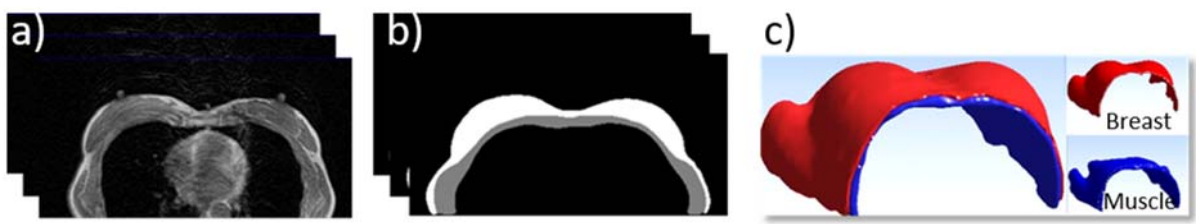

*Fig 1: 3D Geometry generation. a) MR images; b) voxel classification: muscle and breast; c) Corresponding 3D Geometries*

The 3D geometries were meshed using ANSYS Mechanical software[37] with 4-nodes tetrahedra (solid285) in a mixed u-P formulation. Thanks to a sensitivity analysis, the appropriate mesh size was chosen such as the displacement results will be affected by the lower mesh resolution by less than 5%. This constraint resulted in element's sizes ranging between 7 and 10mm. The mesh that was chosen consists in 18453 tetrahedral elements, including 9625 elements that were assigned to the pectoral muscle and the toracic cage, and 8858 elements that were assigned to breast tissue. The element quality was measured using two criteria element: skewness and aspect ratio[37]. The two creteria are ranging between 0.005-0.9 and 1.18-11.53 respectively.

## Stress-free breast configuration

To estimate the stress-free configuration of the breast, an adapted fixed point iterative approach[16] was implemented. Prone and supine image data sets were used to compute the stress-free geometry. The overall iterative process is presented in Fig. 2. At each iteration, the estimated stress-free configuration is used to simulate breast deformation due to gravity in a prone position. The differences between the result of this simulation and the real shape of the breast in prone position is quantified by computing the distance between "active nodes" defined at the breast external surface. For each active node *i*, the point-to-point distance ($D_i$) between its position in the simulated prone configuration and the one in the measured configuration is computed, representing the errors done by the model. This distance is then used in the next iteration of our process to simulate an imposed displacement (Dirichlet condition) to the active node *i* in the stress-free condition. To smooth this imposed displacement and to limit any mesh distortion, the displacement is only partially imposed using a multiplicative regularization factor λ (λ<1). The process repeats as long as the new transformation improves the estimated prone mesh configuration by more than 1mm.

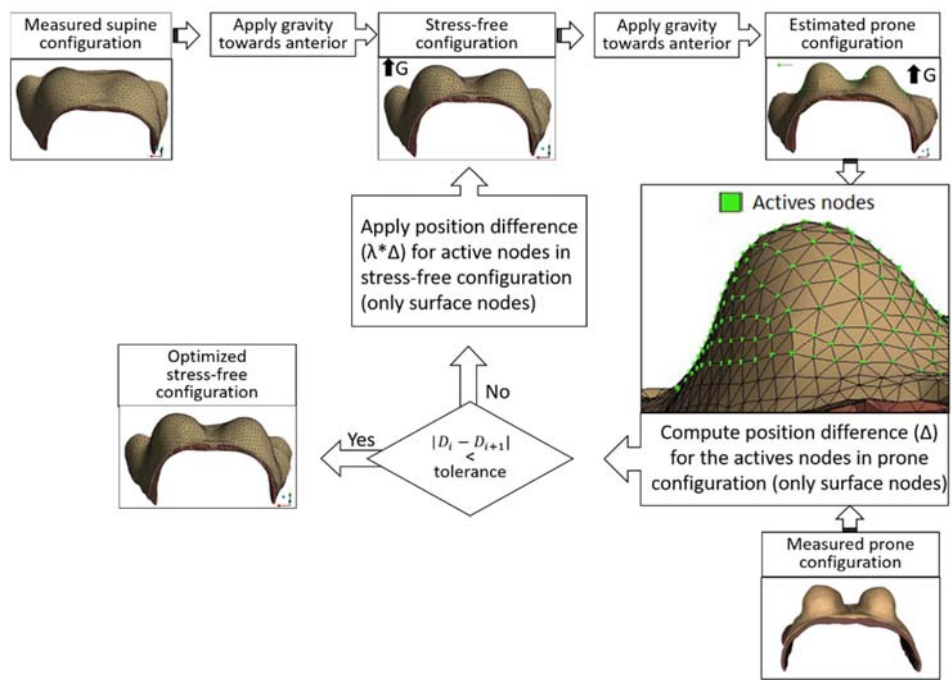

*Fig 2: Fixed point type iterative algorithm for stress-free geometry approximation. Di - mean node to node distance over the active nodes at iteration i, G- gravity force.*

## Boundary conditions

Most recent plastic surgery techniques for breast reconstruction consider the superficial fascial system as the most important component providing support to the breast[38]. Lockwood[38] have investigated the outcome of breast plastic surgery (with a 6 to 36 months follow-up) using a suspension of the superficial fascial system. The results clearly show that this technique provides more stable breast contours and prevents breast "sagging" and "flattening" over time. Therefore, one can conclude that the fascial system has a very important impact on global breast biomechanical behavior and should be included in our model.

The breast fascial system is composed of a deep fascia and a superficial fascia. During puberty, breast is growing and superficial fascia divides in two layers: deep and superficial layers [39]. Cooper's ligaments run throughout the breast. tissue parenchyma from the deep fascia beneath the breast to the superficial layer of superficial fascia where they are attached (Fig. 3.c). Because they are not strained, these ligaments allow the natural motion of the breast [40].

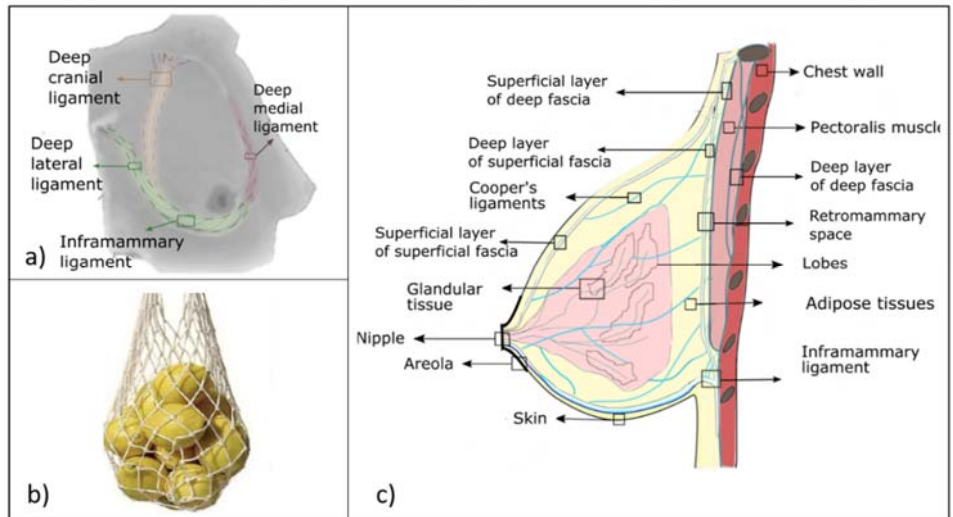

*Fig 3: a) Anatomical position of breast suspensory ligaments. b) Breast support matrix behavior like a filled vegetable mesh bag. c) Breast anatomy.*

Between the superficial layer of the deep fascia and the deep layer of the superficial fascia, a layer of connective loose tissue forms the retro-mammary space, allowing the breast tissue to slide over the chest [41]. In regions where the superficial fascia meets the deep fascia, suspension ligaments are created. The one which is situated at the level of the sixth and seventh ribs is called "inframammary ligament" [42]. The inframammary ligament evolves into the "deep lateral ligament" and the "deep cranial ligament" that are respectively attached to the axillary fascia and to the clavicle (Fig. 3.a). Another suspension ligament is situated on the sternal line (Fig. 3.a) called "deep medial ligament". We strongly believe that the fascial system together with the suspension ligaments create a support matrix for the breast (Fig. 3.b).

As mentioned above, the breast is firmly attached to the deep fascia via suspensory ligaments but moves freely over the pectoralis muscle. We therefore introduced a contact surface between the breast and the muscle implemented as a "no-separation contact" model from ANSYS Contact Technologies. The penalty method is used as a contact algorithm with ANSYS default values to control gap and penetration tolerance factors. The opening stiffness as well as the normal and tangential stiffness factors are adapted for each simulation case in order to ensure the solution convergence [43].

Many simulations with various values of the friction coefficient (k) were performed in order to study its impact on the sliding of the lateral tissues. It was observed that using a different value for k almost does not impact the tissues' lateral displacements when simulating the prone breast configuration. In this work, the deep layer of the superficial fascia is modeled at the interface between the breast and the pectoral muscle. The breast fascia is much stiffer than the breast tissues, therefore the tissues sliding over the pectoral muscle is mainly governed by the fascia's mechanical properties. In order to limit the number of parameters in the model, the friction coefficient was kept constant (k = 0.1) and the amount of sliding was controlled by the fascia's elastic modulus only.

The existence, the topography, and the thickness of the membranous layers of superficial fascia have been studied in various regions of the body by Abu-Hijleh et al. [44]. According to the authors, the thickness of the superficial layer in both superior and inferior breast regions is equal to 88.12 ± 7.70 μm and 140.27 ± 11.03 μm respectively. Therefore, we modeled the deep layer of superficial fascia as a 0.1 mm thick layer of shell elements at the juncture surface between muscle and breast tissue (Fig. 4.1). The superficial layer of superficial fascia is modeled together with the skin as a 2mm thick layer of shell elements over the breast surface. Since the deep fascia and muscle tissues are supposed to present similar elastic properties, we did not explicitly modeled the deep fascia. All ligamentous structures: inframammary ligament, deep medial ligament and lateral ligaments were modeled using Ansys tension-

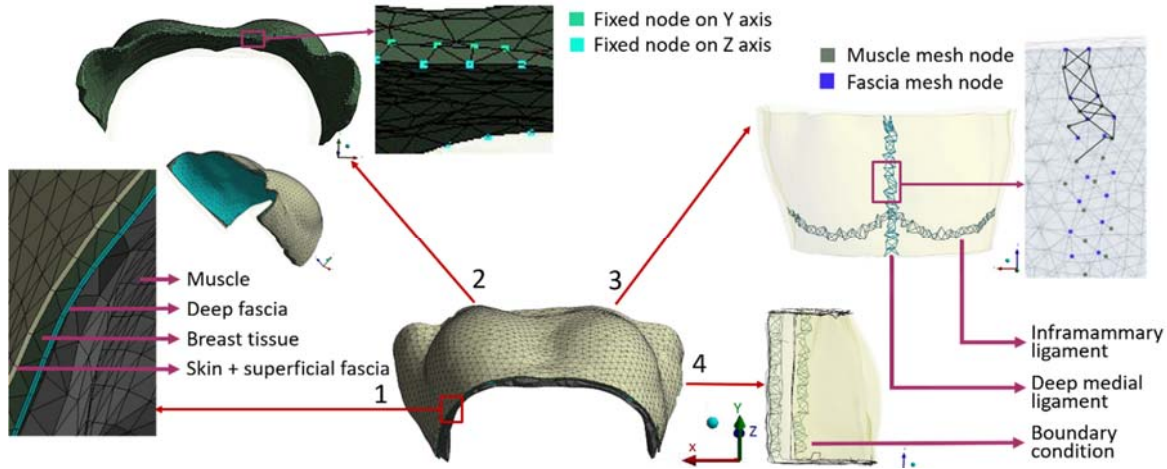
*Fig 4: Finite element mesh components.*

only link elements (LINK180) connecting nodes belonging to the breast mesh component to nodes belonging to the muscle mesh component (Fig. 4.3). Thus, the large lateral displacement derived from breast sliding over the chest wall is mainly limited by fascial and ligamentous connections (Fig. 4.3-4.4).

Dirichlet boundary conditions are added to the posterior part of the pectoral muscle. The superior and inferior ends of the deep fascia layer are constrained in Z direction (Fig. 4.2); the superior and inferior ends of the skin layer are constrained in Y direction (Fig. 4.2). For left and right breast lateral displacements, the Dirichlet conditions are too strong and preclude breast tissue to slide laterally. Therefore, in these regions, such displacements are only constrained by the ligamentous structures with a cable-like behavior (Fig. 4.4).

## Material models

An accurate breast constitutive model stands on a good knowledge of the breast anatomy and on a full characterization of the mechanical behavior for each tissue. The literature proposes models built on four types of tissue: adipose, glandular, muscular tissues and skin (Fig. 2.c). In our work, the breast model is composed of a homogeneous material, and changes with breast granularity. As the left and right breast may have different granularities, they can get different material parameters. Four constitutive models were defined based on previous models (muscle, skin, left and right breast equivalent materials) and two new constitutive models describing the mechanical behavior of fascias and suspensory ligaments: $E_{breast}^r$, $E_{breast}^l$, $E_{muscle}$, $E_{skin}$, $E_{fascia}$, $E_{ligaments}$

Breast soft tissue are considered as nonlinear, anisotropic and time-dependent materials. However, when focusing on breast deformations under gravity loading, the time dependency can be neglected. Assuming that the anisotropic behavior is also negligible, all soft tissue can then be modeled as quasi-incompressible (Poisson's ratio $v$ = 0.49) hyper-elastic Neo-Hookean solids. In that case, the strain-energy density function $W$ is defined by:

$$W = \frac{\mu}{2}(\bar{I}_1 - 3) + \frac{1}{d}(J - 1)^2$$

Where $I_1$ is the first invariant of the left Cauchy-Green deformation tensor, $J$ is the determinant of the deformation gradient F, $\mu$ and $d$ are the initial shear modulus and material incompressibility respectively. For small deformations, $\mu$ and $d$ can be related to Young's modulus ($E$) and Poisson's ratio ($v$) by:

$$\mu = \frac{E}{2(1+v)} \; ; \; d = \frac{6(1-2v)}{E}$$

To our knowledge, there are no published studies describing the mechanical behavior of fibrous structures around the breast, while these structures are well described for legs and arms and are widely included in musculoskeletal biomechanical models. In our work, the anisotropic behavior of the ligaments is neglected and an idealized linear model is used [45,46]. The pectoralis fascia is supposed to have a non-linear stress-strain behavior [47]; therefore it is also modeled as a Neo-Hookean material.

Considering the large variability for the Young's modulus values of breast tissues published in the literature (Table 1), we propose to estimate such moduli for each defined tissue through a sensitivity analysis coupled with an optimization process.

|  | Breast | Reference | Skin | Reference | Fascia | Reference |
|---|---|---|---|---|---|---|
| Min (kPa) | 0.3 | (Gamage et al. 2013)[15] | 7.4 | (Han et al. 2013)[26] | 100 | (Gefen and Dilmoney 2007)[46] |
| Max (kPa) | 6 | (Sinkus 2005)[29] | 58,4 | (Hendriks et al. 2006)[48] | 5000 | (Wenger et al. 2007)[49] |

*Table 1: Minimal and maximal value (in kPa) for equivalent Young's modulus, bibliographic summary.*

### Soft tissue biomechanical properties

Subject-specific mechanical tissue properties were determined using an optimization process based on a multi-gravity loading simulation procedure. First, for a given set of parameters ($E_{breast}^r$, $E_{breast}^l$, $E_{muscle}$, $E_{skin}$, $E_{fascia}$, $E_{ligaments}$), the breast stress-free configuration was estimated by minimizing the difference between the simulated and the measured breast geometries in prone configuration. Then, from the new estimated stress-free geometry, the supine breast configuration was computed and the new estimated geometry was compared to the measured one using modified Hausdorff distance[50]. To exclude the geometry dissimilarity due to subject position, the modified Hausdorff distance was computed only on the breast skin surface.

During the process, multiple simulations based on imposed displacement are performed. Therefore, the FE mesh may be significantly altered before reaching an optimal stress-free geometry. Mainly for that reason, an exhaustive manual search of the optimal set of constitutive parameters was chosen.

Based on existing publications, an interval of possible values was defined for each material (Table 1). As the Young's modulus of breast suspensory ligaments is unknown, it was assumed to be equal to the fascia's Young's modulus. To fine-tune the search intervals of each constitutive parameter, a sensitivity analysis was performed. The results showed that our model is highly sensitive to all previously defined parameters except the Young's modulus of muscular tissues and lateral ligaments. Therefore, the muscle Young's modulus was set to 10 kPa and the laterals ligaments to 100 kPa[49]. It is obvious that beyond some threshold of Young's modulus, materials become too stiff and the breast shape do not change significantly under gravity loadings. Thus, in conformity with the analysis results, the search intervals of Young's moduli were reduced to 0.1-4 kPa, 1-20 kPa and 80-250 kPa for breast, skin and fascia respectively. The new defined intervals were discretized by steps of 0.1 kPa, 1kPa and 40 kPa.

### Results and Discussion

The modified Hausdorff distance between estimated and measured breast geometries in supine configuration is computed at each point of previously discretized intervals. The distance as function of the four independent parameters ($E_{breast}^r$, $E_{breast}^l$, $E_{skin}$, $E_{fascia}$) is shown in Fig 5. The contour lines are estimated by linear interpolation between two consecutive succeeded simulations. For very low values, below $0.2\ kPa, 2\ kPa$ and $80\ kPa$ for breast, skin and fascia's Young's moduli respectively, the tissues deformation is too large and the finite element mesh becomes degenerated at the first step of multi-loading simulation. For values above $1\ kPa$, $5\ kPa$ and $160\ kPa$, tissues deformation is too small compared to the ones measured on the MR images and the simulations were excluded. All other missing values correspond to failed simulations due to a non-converging force, specifically in the region of the contact surface between the breast and the muscle.

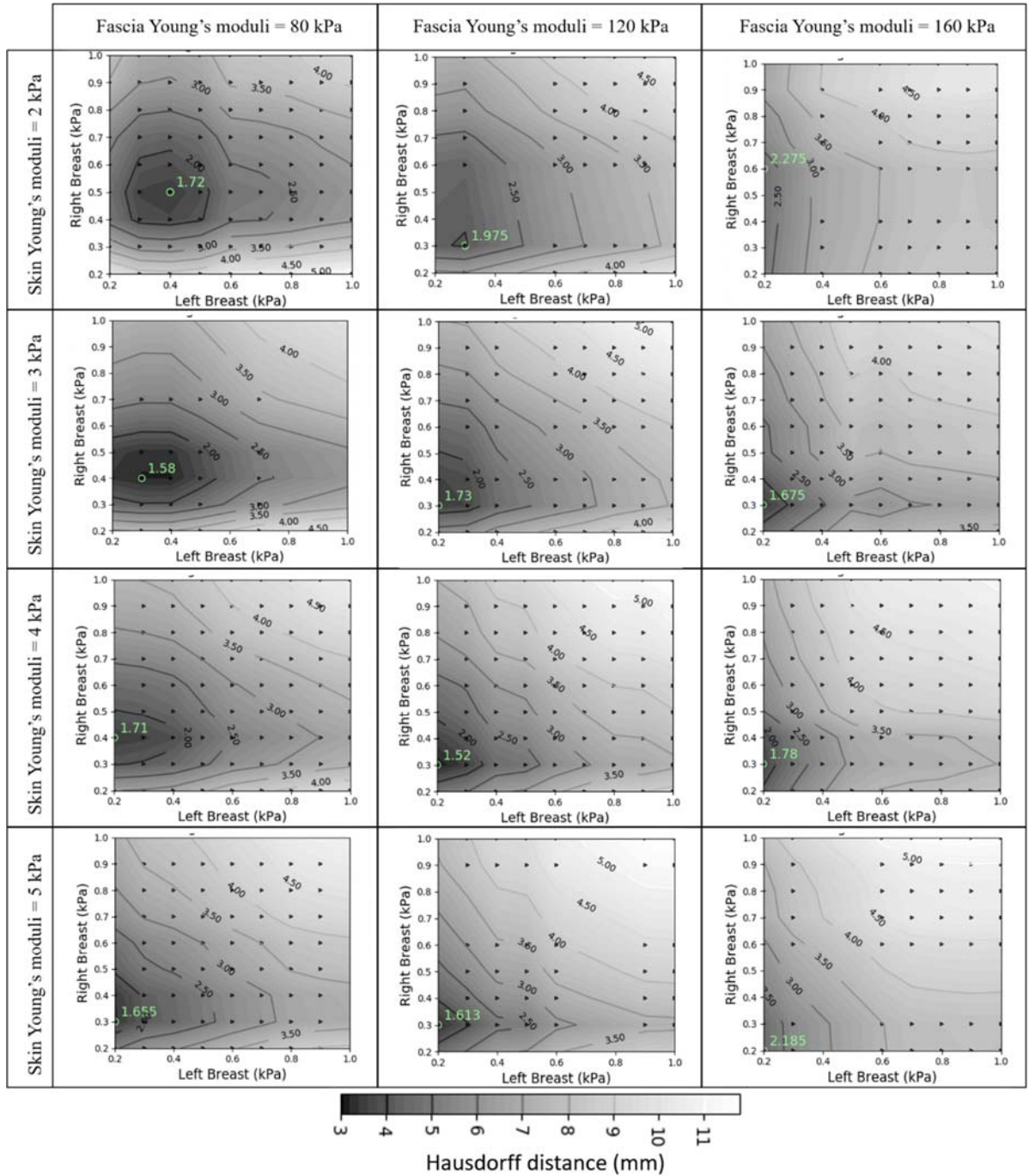

*Fig 5: Hausdorff distance on the skin surface over the constitutive parameters space.*

Fig 6 shows node to surface distance magnitude, mean node to surface distance and modified Hausdorff distance between the simulated and measured breast geometries obtained with the optimal sets of parameters for the three MRI configurations.

The set of parameters giving the best match between simulated and measured supine breast configurations is ($E_{breast}^r = 0.3\ kPa, E_{breast}^l = 0.2\ kPa, E_{skin} = 4\ kPa, E_{fascia} = 120\ kPa$). The breast geometry is better estimated in supine configuration with an Hausdorff distance equal to 1.72 mm. This is probably due to a better representation of the boundary conditions in supine configuration, as this configuration was used to create the initial finite element mesh. The breast geometry in prone configuration is also well estimated with a modified Hausdorff distance equal to 2.17 mm. The maximal node to surface distance is obtained on the breast lateral parts.

One may see that the estimated supine tilted breast configuration describes inadequately the breast geometry given by the MR images (Hausdorff distance equal to 5.90 mm). Many factors may cause such large discrepancies between simulated and measured breast surfaces like an inappropriate definition of the boundary conditions, the lack of pre-tension in skin or glandular tissue, the errors in the estimation of the breast stress-free configuration or the difficulty in the fitting of the materials constitutive models. The estimation of the supine tilted configuration therefore results in abnormally high tissues deformations and excessive tissues sliding over the pectoral muscle. Firstly, such behavior may be caused by an inappropriate definition of the pectoral muscle boundaries. Indeed, neglecting the deformation of the muscle during the repositioning of the volunteer and the possible misalignments during the rigid registration process may impact the estimates in prone and supine tilted breast configurations. Secondly, the fascia and ligamentous tissues are usually characterized by a cable-like behavior. The strain-energy density function must behave asymptotically in order to limit the fascia stretch and thus to reduce non-linearly the breast sliding. The limitations of the Neo-Hookean model to capture the mechanical response of some nonlinear materials is well known[51]. For large strain rates, the Neo-Hookean material may undergo a relaxation and therefore becomes easier to deform. Our experimental results have shown that the maximal strain at the fascia level is significantly higher in supine tilted position (about 140%) than in supine or prone positions (about 50%). Therefore, different material constitutive models considering the asymptotic behavior of fascia mechanical response must be considered. The Gent[52] form of strain-energy function characterizes better such mechanical response and will be tested in the future steps of this work.

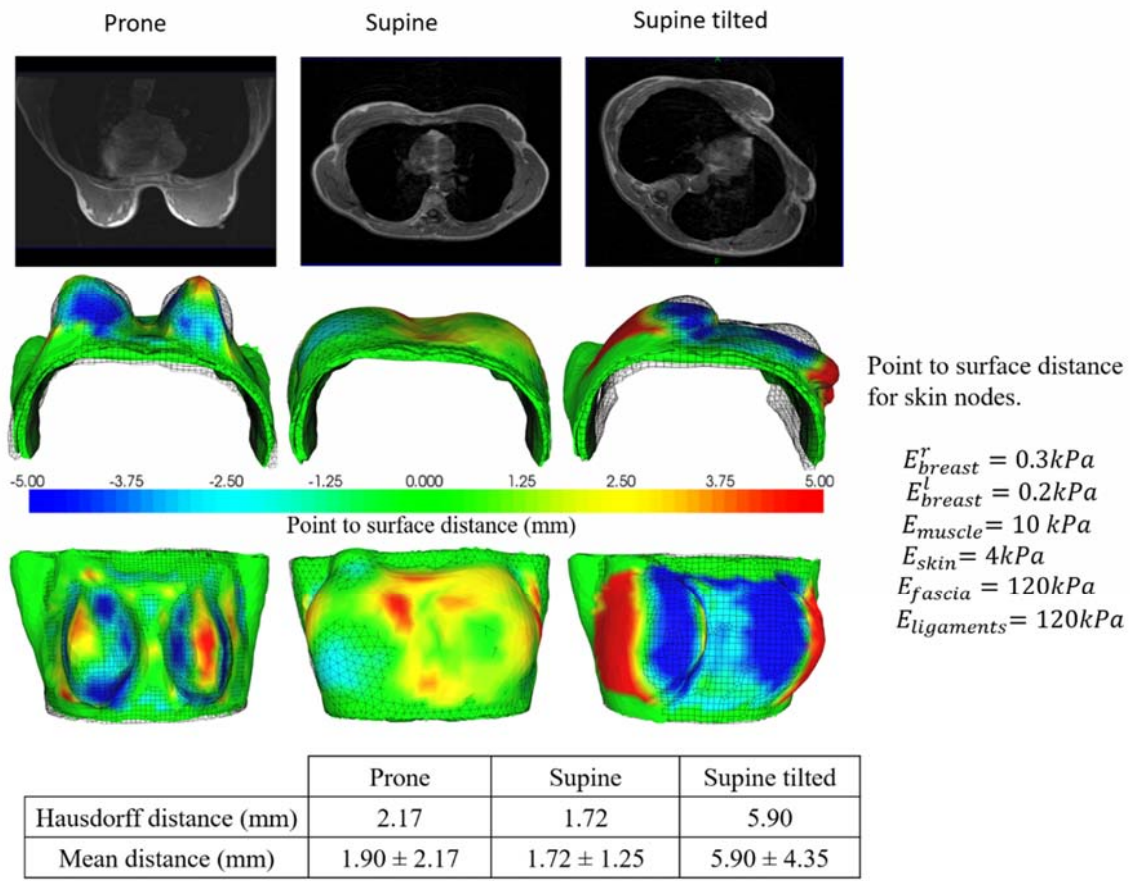

Fig 6: Three breast configurations: prone, supine and supine tilted. First line - MR images in 3 breast configurations. Second and third lines - point to node distance from simulated breast shape (surface mesh) to the measured one (black grid lines).

During the last decades, several breast biomechanical models were proposed ; however, only a small part of them[16,15,26] were evaluated with respect to the real tissues deformations. Gamage and colleagues [15] proposed a finite element model capable to estimate the supine breast configuration from the prone one. To assess the quality of the fit, the root-mean-squared error (RMSE) from point to surface distances was computed. The breast supine geometry was thus estimated within an RMSE of 5mm (maximal distance of 9.3 mm). In the same time, Han and colleagues[26] developed a breast biomechanical model for image registration. The estimates were computed for five subjects, and the accuracy was assessed by computing the Euclidian Distances (ED) between anatomical landmarks. The mean ED ranged between 11.5 mm and 39.2 mm (maximal ED ranged between 20.3mm and 61.7mm). We can thus say that, when evaluated using the predicted skin surface data only, our model shows at least the same performance as the two previous models for the supine and prone estimates. However, for a deeper evaluation of the model, the internal deformations of breast tissues have to be compared with real data. In this scope, internal features within the breast volume such as the distribution of glandular tissues or the spatial location of internal anatomical landmarks have to be estimated and compared with measurements.

## Conclusion

In this work, a new biomechanical breast model was developed using finite element theory. New structures as pectoral fascia and suspensory breast ligaments were considered and their impact on breast mechanics was analyzed in a multi gravity loading simulation. A particular attention was granted to the estimation of subject-specific breast stress-free geometry and tissues constitutive models.

The proposed breast model shows that introducing a sliding movement of the breast tissues over the pectoral muscle together with a ligamentous system, allows a better estimation of supine and prone configurations. Pectoral fascia and breast suspensory ligaments provide a finer method for boundary conditions definition which also improve the convergence capability of the solution. It can be stated that the obtained Young's moduli of breast soft tissues are relatively low (0.2-0.3kPa for breast tissues and 4kPa for skin), which is a contradictory result compared to some studies on the field. However, it is only with such small values together with sliding boundary conditions that prone and supine configurations were accurately estimated.

The model accuracy may be improved by taking into account breast tissue heterogeneity and by introducing subject-specific boundary conditions. As concerns the optimization of the model constitutive parameters, only the distances between the estimated skin nodes and the measured breast surface were considered for the minimization process. However, it is known that the changes of the external shape of the breast cannot entirely describe the breast internal tissues deformations. For a more accurate estimation of the constitutive parameters, the breast internal tissues displacement (via the location of some anatomical landmarks) should be considered. In this work, the supine tilted configuration was only used for evaluation purposes. Since this third configuration can provide a wider set of tissues deformations, it could be interesting to investigate whether including such data into the optimization process will have an impact onto the resulting constitutive parameters.

The major breakthrough of this paper is the model evaluation for three different breast configurations (prone, supine, supine tilted) of the same subject. Moreover, to our knowledge, this is the first time that a biomechanical model is evaluated on a secondary breast configuration which was not used for the model optimization process. The estimate of the supine tilted breast geometry pointed out the limitations of the Neo-Hookean model to assess rich mechanical behavior of breast soft tissues for large strains. These limitations were not identified in the previous works.

For a further usage of the model in the context of breast compression where strains values are very large, another constitutive model will have to be considered. In that perspective, we have recently shown that a Gent constitutive model performs better results than a Neo-Hookean one[53]. Finally, for a better estimation of internal stress during compression, the viscosity of the breast tissue will have to be considered.

## Acknowledgments

This research project is financially supported by ANRT, CIFRE convention n°2014/1357.

We are thanking the IRMaGe MRI facility (Grenoble, France) for their participation in image data acquisition.